\documentstyle[psfig]{mn}

\footnotesize
\newdimen\digitwidth    
\setbox0=\hbox{\rm0}
\digitwidth=\wd0
\catcode`!=\active
\def!{\kern\digitwidth}
\normalsize

\def\msolar{\mbox{M}_{\odot}}
\def\rsolar{\mbox{R}_{\odot}}
\def\pc{\hbox{pc cm$^{-3}$}}
\def\arcdeg{\hbox{$^\circ$}}
\def\arcmin{\hbox{$^\prime$}}
\def\arcsec{\hbox{$^{\prime\prime}$}}

\title{13 Years of Timing of PSR B1259$-$63}

\author[N. Wang et al.]{N.~Wang$^{1,2,3}$, S.~Johnston$^1$ \& R.~N.~Manchester$^2$\\
$^1$ School of Physics, University of Sydney, NSW 2006, Australia \\
$^2$ Australia Telescope National Facility, CSIRO, P.O. Box 76, Epping NSW 1710, Australia \\
$^3$ Urumqi Astronomical Observatory, NAOC-CAS, 40-5 South Beijing Road, Urumqi, China, 830011 \\
}

\begin{document}
\maketitle  

\pagestyle{plain}
\begin{abstract}
This paper summarizes the results of 13 years of timing observations
of a unique binary pulsar, PSR B1259$-$63, which has a massive B2e
star companion. The data span encompasses four complete orbits and
includes the periastron passages in 1990, 1994, 1997 and 2000. Changes
in dispersion measure occurring around the 1994, 1997 and 2000
periastrons are measured and accounted for in the timing
analysis. There is good evidence for a small glitch in the pulsar
period in 1997 August, not long after the 1997 periastron, and a
significant frequency second derivative indicating timing noise.  We
find that spin-orbit coupling with secular changes in periastron
longitude and projected semi-major axis ($x$) cannot account for the
observed period variations over the whole data set. While fitting the
data fairly well, changes in pulsar period parameters at each
periastron seem ruled out both by X-ray observations and by the large
apparent changes in pulsar frequency derivative. Essentially all of
the systematic period variations are accounted for by a model
consisting of the 1997 August glitch and step changes in $x$ at each
periastron.  These changes must be due to changes in the orbit
inclination, but we can find no plausible mechanism to account for
them. It is possible that timing noise may mask the actual changes in
orbital parameters at each periastron, but the good fit to the data of
the $x$ step-change model suggests that short-term timing
noise is not significant.
\end{abstract}

\begin{keywords}
Binaries: general -- pulsars: individual: PSR B1259$-$63
\end{keywords}

\section{Introduction}
PSR B1259$-$63 was discovered in a large-scale high frequency survey of
the Galactic plane \cite{jlm+92}. It is unique because it is
the only known radio pulsar in orbit about a massive, 
main-sequence, B2e star \cite{jml+92}. PSR B1259$-$63 has a
short spin period of $\sim$48~ms and moderate period derivative of
$2.28\times10^{-15}$, implying a characteristic age of only 330~kyr
from spin-down by magnetic dipole radiation. It has an orbital period
of $\sim$1237 days, and an eccentricity of 0.87, the longest orbital
period and largest eccentricity of any of the known binary radio pulsars.
The inclination of the orbit to the plane of the sky is $i\sim36\arcdeg$.

The companion, SS 2883, is a 10th magnitude star with a mass of about
10~$\msolar$ and a radius of 6~$\rsolar$. Typical of Be stars as a
class, it has a hot, tenuous polar wind and a cooler, high density,
equatorial disk. The mass loss rate of the B2e star is $\sim
10^{-6}$~$\msolar$.  Johnston et al. (1994)\nocite{jml+94} observed
H$\alpha$ emission lines from the disk at 20 stellar radii (R$_{*}$),
just inside the pulsar orbit of 24 R$_{*}$ at periastron. The density
of the disk material is high near the stellar surface
($10^{8}-10^{10}$~cm$^{-3}$), and falls off as a power-law with
distance from the star. The disk is likely to be highly tilted with
respect to the pulsar orbital plane, and PSR B1259$-$63 is eclipsed for
about 40 days as it goes behind the disk.

Assuming the B2e star is rotating at $\sim$70 per cent of its break up
velocity \cite{por96}, SS 2883 then has an equatorial velocity
$\sim$280~kms$^{-1}$.  The spin-induced oblateness of the star implies
an additional $1/r^3$ gravitational potential term in the interaction
with the pulsar, known as quadrupole gravitational moment in typical
binary system. This effect introduces an apsidal motion and precession
of the orbital plane if the spin of the companion is not aligned with
the orbit angular momentum, characterised by $\dot \omega$ and $\dot
x$ respectively, where $\omega$ is the longitude of periastron passage
and $x$ is the projected pulsar semi-major axis \cite{lbk95}. 

In an eccentric binary system, the passage of the pulsar through
periastron excites tidal motions on the companion star which then
interact with the orbital motion. In the case of PSR J0045$-$7319
\cite{kbm+96}, tidal effects may account for the observed evolution in
binary parameters (e.g. Lai 1997)\nocite{lai97}. The tidal interactions
may be enhanced by large factors if there is a resonance with
oscillation modes of the star \cite{ws99}. For PSR J0045$-$7319, at
periastron the pulsar is only $\sim 4$ R$_{*}$ from the companion
compared to 24 R$_{*}$ for PSR B1259$-$63. Tidal effects are strongly
dependent on the separation of the two stars, and so 
may be insignificant in PSR B1259$-$63. 


Interactions between the pulsar and the disk of the Be star may also
be important. If there is accretion from the disk on to the neutron
star, the pulsar may be spun up or slowed down. For accretion to
occur, the bow shock, where there is pressure balance between the
pulsar wind and the Be-star disk gas, must lie inside the accretion
radius, which depends on relative velocity of the pulsar and the
accreting gas. For PSR B1259$-$63, accretion is unlikely
\cite{mjl+95,ktn+95,ta97}, with X-ray observations \cite{hck+99} most
easily interpreted as originating from a bow shock well outside the
pulsar magnetosphere. Disk interactions may also affect pulsar orbit
directly through frictional drag.  Estimates \cite{mjl+95,wjm+98} of
variations in the Keplerian parameters $P_b$, $\omega$ and $e$ suggest the
effect is negligible.

Timing models for binary pulsars have evolved over the years, mainly
to take into account general relativistic effects \cite{bt76,dd86}.
Wex (1998)\nocite{wex98} extended these models to deal with
long-period systems with main sequence companions. His model, the
so-called MSS model, accounts for short-term periodic orbital effects
and for the long-term secular effects caused by classical spin-orbit
coupling. 

The latest paper on timing of PSR~B1259$-$63 \cite{wjm+98} analysed
observations from 1990 January to 1996 December which covered the 1990
and 1994 periastron passages. Wex et al. (1998) were unable to derive
a unique timing model for the system, partly because of the large
timing noise and partly due to the long eclipses of the pulsed
emission.  They put forward two possible solutions for the timing. In
one, timing noise dominated, its effects were removed through higher
order derivatives of the spin frequency, and there was no orbital precession or
tidal effects.  In the second model, which they preferred, significant
values of $\dot \omega$ and $\dot x$ were obtained. These were
interpreted as being due to orbital precession, and a large tilt
between the rotation axis of the Be star and the orbital angular
momentum vector was inferred. Two different fits
were obtained for $\dot \omega$ and $\dot x$ depending on the number
of (integer) rotations added during the 1994 eclipse. The preferred
solution then relied on the physical interpretation of the values
of $\dot \omega$ and $\dot x$. 

Since then, we have obtained a further six years of timing data on
this pulsar to give a total data span of 13 years, including two
further periastron passages in 1997 and 2000.  Our main findings are:
(1) that there is strong evidence for a pulsar period glitch in 1997
August, about 94 days after the 1997 periastron, (2) that the Wex et
al. (1998) timing solution which included orbital precession does not
remain valid through the 1997 and 2000 periastrons, (3) that a
solution with jumps in $\nu$ and $\dot \nu$ at each periastron, while
fitting the data reasonably well, is unlikely to be physically viable,
and (4) that the data are best explained by a combination of the 1997
August pulsar glitch and steps in projected orbit semi-major axis
($x$) at each periastron.  However, the possible presence of timing
noise and the long-duration eclipses conspire to make timing of the
pulsar difficult. In Section~\ref{sec:obs} we discuss details of the
observations and data analysis techniques, Section~\ref{sec:prf}
describes the pulse profile evolution from 0.66 to 13.6~GHz and
Section~\ref{sec:ddm} describes the dispersion measure variations
observed around the 1994, 1997 and 2000 periastrons. The timing
results are described in Section~\ref{sec:tim} with fits of the
various possible models, and the possible interpretations are
discussed in Section~\ref{sec:dis}.

\section{Observations and Analysis}\label{sec:obs}
We report here on observations of PSR B1259$-$63 using the Parkes radio
telescope over a time interval of nearly 5000 days between 1990
January and 2003 June (MJD 47909 to 52804). A total of 1031 times of
arrival (TOAs), spanning four orbital periods, were obtained for the
pulsar.  The majority of the observations were made at frequencies
around 1.4~GHz with additional observations at 0.66, 2.4, 4.8, 8.4
and 13.6~GHz.

At 1.4~GHz, the H-OH receiver \cite{tgj90} with a system equivalent
flux density of $\sim$40~Jy was used prior to 1997. From 1997, the
multibeam receiver \cite{swb+96}, with a system equivalent flux
density of $\sim$28~Jy, was employed. Short observations with duration
$\sim$10~min were made at typically two-week intervals far from
periastron. Longer observations with duration $\sim$1~hr were made on
a daily basis at times close to periastron. The higher frequency
observations were generally conducted close to the epochs of
periastron in order to measure the dispersion measure changes and the
scatter broadening of the pulse profile \cite{jwn+01}. The receivers
at 4.8 and 8.4~GHz had system equivalent flux densities of
$\sim$100~Jy.  Lower frequency observations were more rarely
performed. The combination of increased Galactic background emission
at low frequencies, and the flat spectral index of the pulsar, made it
hard to detect.

The backend systems included various combinations of filterbanks and
correlators. Table~\ref{tb:system} lists the details of the observing
systems. Filterbank systems were used throughout the observation
period. At frequencies above 1~GHz, they generally consisted of 64
frequency channels each of width 5~MHz (1991-1997), 96 channels of
width 3~MHz (1997-203) or 512 channels of width 0.5~MHz (2001
onwards). The filterbank backends were only capable of measuring total
power. They one-bit sampled the data typically at intervals of
600~$\mu$s for this pulsar. The data stream was written to tape for
subsequent off-line analysis. This analysis involved de-dispersing and
folding the data at the topocentric period.  A digital correlator, the
Caltech Fast Pulsar Timing Machine (FPTM; Navarro 1994\nocite{nav94}),
was in use between 1994 and 2001. It had 128~MHz of bandwidth with 128
frequency channels. When operating in total power mode, two
independent 128~MHz bands were available. When full Stokes parameters
were recorded, however, only one band could be used.  The correlator
performed on-line pulse-folding and dedispersion to produce a pulse
profile with up to 1024 time bins per period.  Further information for
earlier observations can also be found in Wex et al. (1998) and
Johnston et al. (2001).

\begin{table}
\caption{Detail of the observing systems used for PSR B1259$-$63. FB denotes
a filterbank system  and FPTM denotes the Caltech Fast Pulsar Timing Machine.}
\begin{tabular}{lclccc}
\hline    & \vspace{-3mm} \\
Freq &  Backend  & \multicolumn{1}{c}{Year} &\multicolumn{1}{c}{BW}  &  \multicolumn{1}{c}{No. of}    &\multicolumn{1}{c}{No. of} \\
\multicolumn{1}{c}{(GHz)}&           &                          &\multicolumn{1}{c}{(MHz)}&\multicolumn{1}{c}{Channels}          &\multicolumn{1}{c}{TOAs}    \\   
\hline  & \vspace{-3mm} \\
0.66    &  FB     & 1990--1996       & ~~32 & 128        &  ~~14  \\
        &  FPTM   & 1995--1997       & ~~32 & 128        &  ~~13  \\
1.4     &  FB     & 1990--1997       & 320  & ~~64       &  243 \\
        &  FB     & 1997--2003       & 288  & ~~96       &  101  \\
        &  FB     & 2001             & 256  & 512        &  ~~~3    \\ 
        &  FPTM   & 1994--2001       & 128  & 128        &  444 \\
        &  FPTM   & 1997--1998       & ~~64 & 128        &  ~~~4  \\
2.4     &  FB     & 1992--1993       & 320  & ~~64       &  ~~11\\
        &  FPTM   & 2000--2001       & 128  & 128        &  ~~~4 \\
4.8     &  FB     & 1993--1997       & 320  & ~~64       &  ~~33 \\
        &  FB     & 1997             & 576  & 192        &  ~~24 \\
        &  FPTM   & 1996--2001       & 128  & 128        &  ~~78 \\
8.4     &  FB     & 1993--1996       & 320  & ~~64       &  ~~21  \\
        &  FB     & 1997             & 576  & 192        &  ~~~5  \\
        &  FPTM   & 1996--1997       & 128  & 128        &  ~~17  \\ 
13.6    &  FPTM   & 1995             & 128  & 128        & ~~15 \\ 
\hline  & \vspace{-3mm} \\
\end{tabular}
\label{tb:system}
\end{table}          

\begin{figure}
\centerline{\psfig{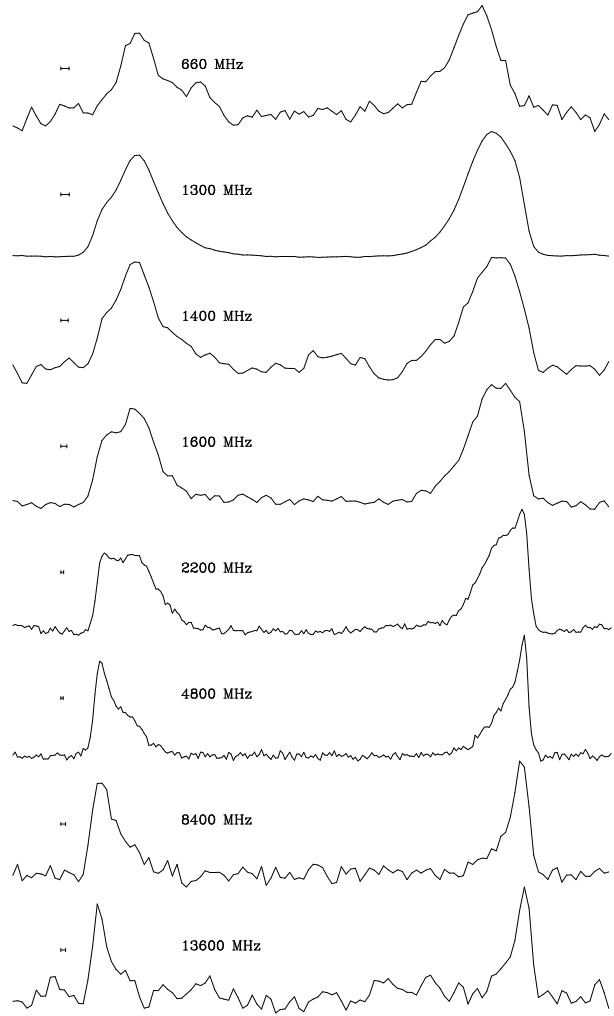}}
\caption{Standard profiles of PSR B1259$-$63, summed from individual
observations. The bar on the left side of the profiles is the
effective resolution, including the effects of DM smearing.}
\label{fg:std}
\end{figure}

The pulse profile of PSR B1259$-$63 consists of two, almost
equal-strength components.  Each component has a steep edge and a more
slowly falling edge.  Manchester \& Johnston (1995) \nocite{mj95}
suggested that the steep edges were the outside of a cone with large
opening angle. In order to obtain a good value for the dispersion
measure, it is important that the profile be properly aligned over the
entire frequency range between 0.66 and 13.6~GHz. We chose the point
midway between the mid-points of the steep outer edges as the fiducial
point.  At each frequency, a large number of profiles were summed
together to form a high signal-to-noise template. These templates are
shown in Fig.\ref{fg:std}. To obtain an accurate TOA we then
cross-correlated individual observations with the relevant template.

Table~\ref{tb:gap} lists the epoch of periastron for each of the
four orbits so far observed. It lists the total number of TOAs obtained
during the orbit, and indicates the first detection of the pulsar after
periastron and the final detection of the pulsar before it enters the eclipse.

\begin{table}
\caption{Details of the observation data spans. 
$\tau+$ is the first detection of the pulsar after the preceding periastron and
$\tau-$ is the final detection prior to the following periastron.}
\begin{tabular}{llrrr}
\hline         & \vspace{-3mm} \\
   Data         &  Preceding   & $\tau+$     & $\tau-$ & No. of \\
   span         &  Periastron (MJD) &  (d)       & (d)    & TOAs \\ 
\hline         & \vspace{-3mm} \\
1990.1--1990.7 &     --        &  --         & 107     &  18  \\
1990.7--1994.0 &    48124      &  171        & 20      &  187 \\
1994.0--1997.5 &    49361      &  24         & 18      &  443 \\
1997.5--2000.9 &    50597      &  16         & 52      &  237 \\   
2000.9--2003.5 &    51834      &  19         & --      &  146 \\   
\hline & \vspace{-3mm} \\
\end{tabular}
\label{tb:gap}
\end{table}          

The eclipses of the pulsar, which last for $\sim$40 days around
periastron make the timing difficult.  The problem is exacerbated by
the fact that DM changes are observed over a few tens of days leading
up to and immediately following the eclipse due to the changing line
of sight intersecting the Be star wind.
These DM variations need to be removed to obtain a good timing
solution, and this involves making observations at several different
frequencies as near to simultaneously as possible.
Multi-frequency observations were therefore made before and after the 1994,
1997 and 2000 periastron passages and have been described in detail in
Johnston et al. (1996) and Johnston et al. (2001).  Generally,
observations were made at 5 different frequencies, 1.2, 1.4, 1.5, 4.8
and 8.4~GHz. The three frequencies near 1.4~GHz were obtained
simultaneously and the observation would then be followed by one at
either 4.8 or 8.4~GHz before again observing at the lower
frequencies. 

In order to measure an accurate DM we assume that the DM
contribution from the 1.5~kpc along the line of sight through the
interstellar medium is 146.8~pc~cm$^{-3}$ \cite{cjmm02}.  Then the
extra DM contribution from the disk of the Be star could be computed
from the relative difference in the timing residuals between the
different frequencies. Typical errors using this method are
$\sim0.1$~$\pc$.

The timing properties were analysed using the pulsar timing program
TEMPO\footnote{See
http://www.atnf.csiro.au/research/pulsar/timing/tempo/}, which
provides least-squares fitting to the pulsar rotation and orbital
parameters and the dispersion measure. The MSS binary model of Wex (1998)
was used for all fits except those in Section~\ref{sec:btj}. To fit
for steps in the orbital parameters, a new binary model (BTJ) based on
the BT model \cite{bt76} was implemented. This allows cumulative steps in
longitude of periastron ($\omega$), projected semi-major axis ($x$),
eccentricity ($e$) and binary period ($P_b$) to be inserted at
specified times and also allowed setting or solving for jumps in
pulsar phase at the specified times. 

We note that our results differ slightly from those of Wex et
al. (1998) when analysing the same data set. This is because new
standard profile templates were used and the dispersion analysis
re-done for this work.

\section{Pulse Profile Evolution}\label{sec:prf}
Fig.~\ref{fg:std} shows that the pulse profile evolves substantially
over the observed range from 660 to 13600 MHz. At first glance it
appears that the profile gets narrower (when plotted as in
Fig.~\ref{fg:std}) at lower frequencies, the opposite trend to that
observed in most conal profiles. However, a closer examination
suggests that this apparent width variation is largely due to
differing spectral index of components making up each peak of the
profile. The outer components are stronger at higher frequencies
whereas components on the inner edge of each peak become strong at
lower frequencies and dominate the profile at frequencies below
2~GHz. The flatter spectral index for outer conal components is
commonly observed in other conal pulsars \cite{ran83,lm88} and
reinforces the interpretation that the steep edges define the boundary
of a wide emission cone from a single polar region (cf. Manchester
1996\nocite{man96}).

\section{DM Variations}\label{sec:ddm}
\begin{center}
\begin{figure}
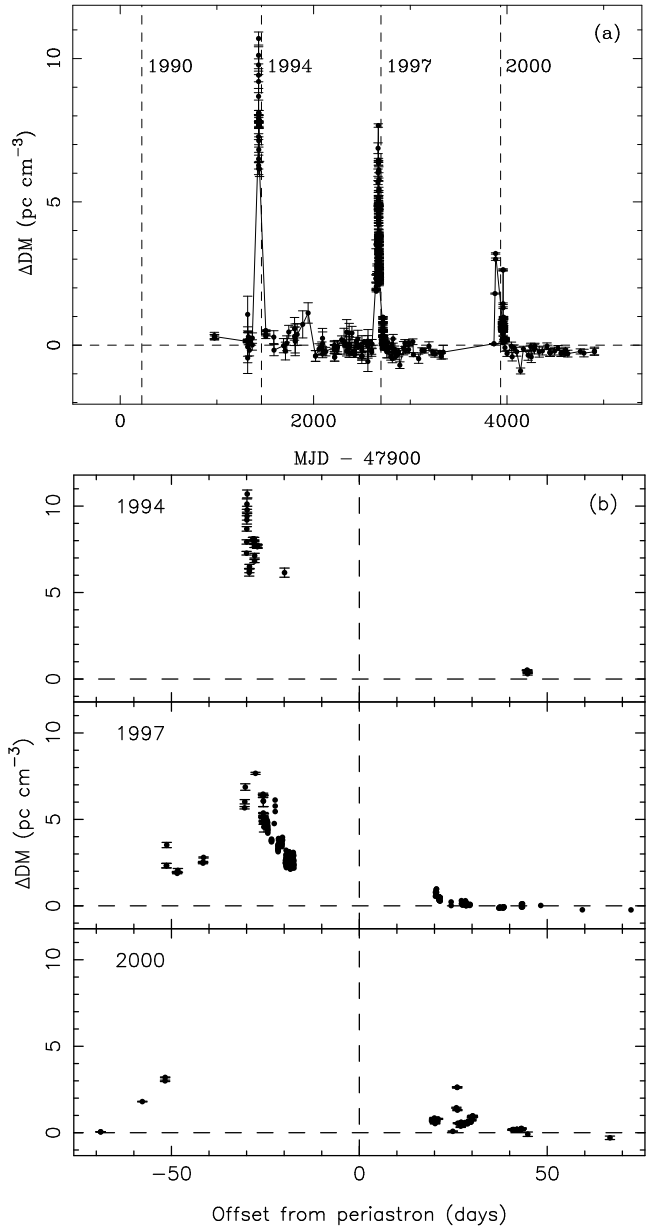
 
\begin{tabular}{c}
\mbox{\psfig{file=fig2a.ps,width=84mm,angle=270}}\\
\mbox{\psfig{file=fig2b.ps,width=85mm}}\\
\end{tabular}
\caption{(a) Dispersion measure variations from 13 years of timing
observation.  The four vertical lines in this and subsequent plots
represent 1990, 1994, 1997 and 2000 periastrons for PSR B1259$-$63. (b)
Expanded plots for DM variations $\pm$75 days around periastron. The offset
is measured relative to the value of 146.8~$\pc$. }
\label{fg:dm}
\end{figure}
\end{center}

As PSR B1259$-$63 approaches periastron and moves into the Be-star
disk, electron density variations along the line of sight result in DM
changes as a function of orbital phase. Fig.~\ref{fg:dm}a shows the
observed variations from 1992, covering the 1994, 1997 and 2000
periastrons, and Fig.~\ref{fg:dm}b shows expanded plots for $\pm75$
days about the periastrons.  DM variations for the 1994 periastron
were reported by Johnston et al. (1996)\nocite{jml+96}, 
and for the 1997 periastron these were monitored extensively and
reported by Johnston et al. (2001), while the 2000 variations
have not been previously reported.

At the 1994 periastron, a peak of $10.7\pm0.2$~$\pc$ for $\Delta$DM
was observed at $\tau-30$; after this it decreased with some
fluctuations until the pulsar was eclipsed.  In 1997, the peak was
$7.7$~$\pc$ at $\sim\tau-28$, slightly later than in 1994, and again
the $\Delta$DM drops with some fluctuations until the pulsar is
eclipsed at $\sim\tau-18$.  There are no multi-frequency data prior to
the 2000 periastron.  The points of $\Delta$DM in Fig.\ref{fg:dm} are
derived by extrapolating a timing solution based on observations in
the 60 days before these pre-periastron points shown in
Fig.~\ref{fg:dm}b. The latest pre-periastron point is 52 days before
the periastron. Based on the 1994 and 1997 behaviour, the peak of
$\Delta$DM was not observed.

The $\Delta$DM fluctuations are well observed after the 1997 and 2000
periastrons. The $\Delta$DM values are much smaller than before
periastron and approach zero about 30 days after periastron. In fact,
Fig.~\ref{fg:dm}~(a) shows that the $\Delta$DM values are slightly
negative in mid-orbit, suggesting the reference DM value of
146.8~$\pc$ is too large. The filterbank observations from $\tau+43$
after the 2000 periastron to the end of the data set gives a mean
offset of $-0.2\pm0.1$~$\pc$ from the reference value; corresponding
to an interstellar dispersion of 146.6$\pm$0.1~$\pc$.

\section{Timing Results}\label{sec:tim}
\subsection{Simulations}\label{sec:sim}
\begin{figure} 
\centerline{\psfig{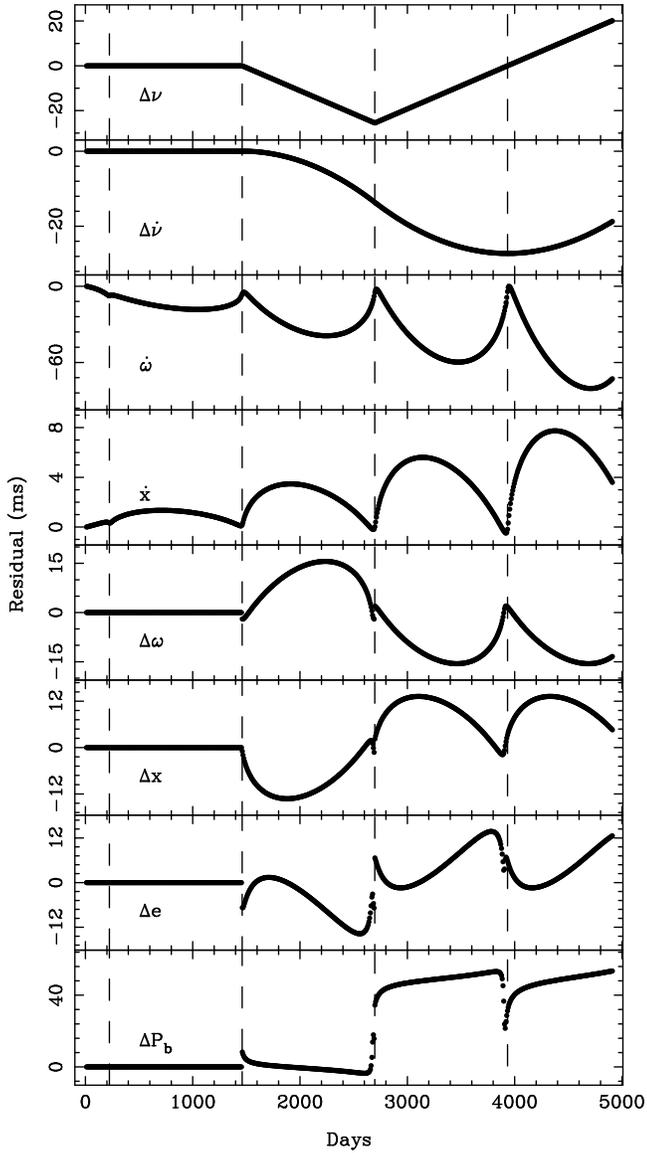}}
\caption{Simulated timing residuals after including jumps in $\nu$ and
$\dot{\nu}$ (top two panels), constant $\dot{\omega}$ and $\dot{x}$
(next two panels) and jumps in orbital parameters (bottom four
panels). Jumps were inserted at the second and third periastrons. 
See text for details.}
\label{fg:many}
\end{figure}

We first used simulations to determine the shape and amplitude of
timing residuals corresponding to various parameters used in the
fitting procedure. Fig.~\ref{fg:many} shows us the timing signatures
of the various parameters and the size of the resultant phase
perturbations.  In the figure we have attempted to ensure that the
resultant residuals were of the order of tens of milliseconds, similar
to the residuals seen in the real data. Apart from the $\dot{\omega}$
and $\dot{x}$ simulations, we introduced steps at the second and third
periastron passages, with the change at the third periastron being
twice the magnitude and of opposite sign to the change at the second
periastron. Since the changes are cumulative, this has the effect of
reversing the sign of the perturbation at the third periastron. The
steps in $\nu$ and $\dot\nu$ at the second periastron are respectively
$5\times 10^{-9}$~s$^{-1}$ and $5\times 10^{-17}$~s$^{-2}$. For
$\dot{\omega}$ and $\dot{x}$ we use the values from the Wex et
al. (1998) model 2A of $2\times 10^{-4}$~deg~yr$^{-1}$ and $15\times
10^{-12}$ respectively.  The orbital parameter steps at the second
periastron are $\Delta\omega = 0\fdg03 $, $\Delta~x = 0.01$~s, $\Delta
e = 4\times 10^{-6}$ and $\Delta P_b = 30$~s respectively.

\subsection{A Glitch Near MJD 50691}
After we had been through the process of fitting the data as described
in the subsections below, it became apparent to us that there was
strong evidence for a glitch in the pulsar period near MJD 50691 (1997
August 30), about 94 days after the 1997 periastron.  The evidence for
the glitch and its parameters are described more fully in
Section~\ref{sec:btj} below. In each of the following subsections we
describe the fits both with and without the glitch. In
all cases except the final fit in Section~\ref{sec:btj}, the glitch
parameters are held fixed at the values determined in this final fit.

\subsection{The Keplerian Timing Model}\label{sec:kep}
\begin{figure} 
\centerline{\psfig{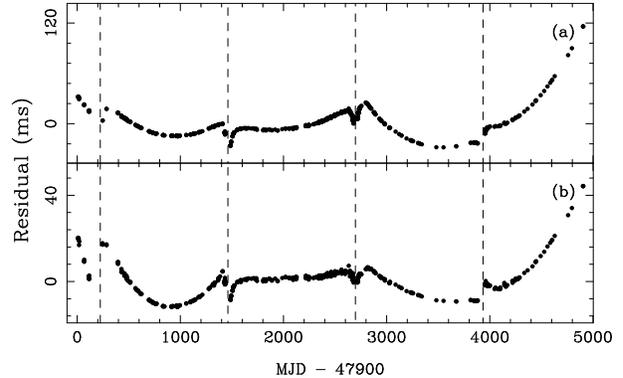}}
\caption{(a) Post-fit residuals after fitting for $\nu$, $\dot\nu$,
$\ddot\nu$ and the Keplerian orbital parameters.  The rms residual is
14.6~ms. (b) Post-fit residuals with the 1997 August glitch
(Section~\ref{sec:btj}) included, giving an rms residual is 4.3~ms.}
\label{fg:nowx}
\end{figure}

We initially attempted to fit the entire data set using the five
Keplerian orbital parameters and the spin frequency and its first two
time-derivatives. Wex et al. (1998) already showed that this did not
fit the data well. The residuals for our extended data set are shown
in Fig.~\ref{fg:nowx}a where the rms residual is 14.6~ms.  The
post-fit residuals after including the glitch shown in
Fig.~\ref{fg:nowx}b are somewhat reduced with an rms residual of
4.3~ms, but still have large systematic variations. Parameters from
this fit are given in the second column of Table~\ref{tb:para}. If we
add four more higher-order spin frequency derivatives, the rms
residual reduces to $\sim 2.1$~ms but only at the expense of
significantly more free parameters and still leaving systematic
residuals. We note that in the timing residual plots
(Fig.~\ref{fg:nowx} -- \ref{fg:obj}) residual error bars are not
plotted since the TOA errors (typically $<$100~$\mu$s) are generally
smaller than the plotted points. In general, systematic departures
from the timing model dominate the residual plots.

\begin{table*}
\caption{Four timing solutions for PSR B1259$-$63.}
\begin{tabular}{lllll}
\hline & \vspace{-3mm} \\
                             &\multicolumn{1}{c}{No $\dot\omega$, $\dot x$}    
                             &\multicolumn{1}{c}{$\dot\omega$, $\dot x$}    
                             &\multicolumn{1}{c}{$\nu$, $\dot\nu$ jumps} 
                             &\multicolumn{1}{c}{Orbital jumps}   \\
\hline & \vspace{-3mm} \\
R.A. (J2000)                 &  13$^{h}$02$^{m}$47$^{s}$.65(1)
                             &  13$^{h}$02$^{m}$47$^{s}$.65(1)
                             &  13$^{h}$02$^{m}$47$^{s}$.65(1)
                             &  13$^{h}$02$^{m}$47$^{s}$.65(1)\\
Dec. (J2000)                 &  $-$63$\arcdeg$50$\arcmin$08$\arcsec$.7(1) 
                             &  $-$63$\arcdeg$50$\arcmin$08$\arcsec$.7(1) 
                             &  $-$63$\arcdeg$50$\arcmin$08$\arcsec$.7(1) 
                             &  $-$63$\arcdeg$50$\arcmin$08$\arcsec$.7(1)\\
DM ($\pc$)                   &  146.8 &  146.8 &  146.8 &  146.8         \\
$\nu$ (s$^{-1}$)             & 20.93692453667(7)  & 20.9369245383(1) & 20.93692435(4) &20.9369245339(9)\\
$\dot \nu$ $(\times 10^{-12}$ s$^{-2})$ & $-$0.9979077(6)  & $-$0.9978663(7)  & $-$0.9987(2)& $-$0.9979383(9) \\
$\ddot\nu$ $(\times 10^{-24}$ s$^{-3})$ & $-$2.11(2)       & $-$1.96(2)      & 0.5(5)     & $-$1.762(7)  \\
Period epoch (MJD)             &  50357.00   &  50357.00   &  50357.00   &  50357.00  \\
$x$ (s)                      & 1296.315(2)  & 1296.282(2)    & 1296.3264(4)  & 1296.272(5) \\
$e$                          & 0.8698869(5) & 0.8698832(6)   & 0.8698902(2)  & 0.8698872(9)\\
$t_0$ (MJD)                  & 48124.3491(1)& 48124.3494(2)  & 48124.34892(3)& 48124.34911(9)\\
$P_b$ (d)                    & 1236.72404(3)& 1236.72360(7)  & 1236.724259(7)& 1236.72432(2)\\
$\omega$ (deg)               & 138.66588(7) & 138.6644(1)    & 138.66624(6)  & 138.6659(1)\\
Glitch epoch (MJD)             &  50690.7 &  50690.7 &  50690.7 &  50690.7(7)\\
$\Delta\nu_g$ $(\times 10^{-9}$ s$^{-1})$  &67     &  67      &  67     & 67(1)\\
$\Delta\nu_d$ $(\times 10^{-9}$ s$^{-1})$  &22     &  22      &  22     & 22(1)\\
$\dot \omega$ (deg yr$^{-1}$)               &$-$    &0.00020(1) & $-$       & $-$ \\
$\dot x$ ($\times 10^{-12}$)                       &$-$    & 127(5)    & $-$       & $-$  \\
$\Delta\nu_{90}$ $(\times 10^{-9}$ s$^{-1})$         & $-$   & $-$    &  0(3)     &  $-$\\
$\Delta\dot \nu_{90}$ $(\times 10^{-15}$ s$^{-2})$   & $-$   & $-$    &  0.8(2)      &  $-$\\ 
$\Delta\nu_{94}$ $(\times 10^{-9}$ s$^{-1})$         & $-$   & $-$    &  15(3)       &  $-$ \\
$\Delta\dot \nu_{94}$ $(\times 10^{-15}$ s$^{-2})$   & $-$   & $-$    &  0.19(5)     &  $-$\\
$\Delta\nu_{97}$ $(\times 10^{-9}$ s$^{-1})$         & $-$   & $-$    &  $-$10(3)    &  $-$\\
$\Delta\dot \nu_{97}$ $(\times 10^{-15}$ s$^{-2})$   & $-$   & $-$    &  $-$0.30(5)  &  $-$\\
$\Delta\nu_{00}$ $(\times 10^{-9}$ s$^{-1})$         & $-$   & $-$    &  $-$2(3)  &  $-$\\
$\Delta\dot \nu_{00}$ $(\times 10^{-15}$ s$^{-2})$   & $-$   & $-$    &  $-$0.33(5)  &  $-$\\
$\Delta x_{90}$ (ms)               & $-$   &  $-$   &   $-$   &  60.3(7)\\   
$\Delta x_{94}$ (ms)               & $-$   &  $-$   &   $-$   &  $-$26.3(1)\\   
$\Delta x_{97}$ (ms)               & $-$   &  $-$   &   $-$   &  2.8(3) \\   
$\Delta x_{00}$ (ms)               & $-$   &  $-$   &   $-$   &  4.2(4) \\   
No. TOAs                           & 1031  &  1031  &  1031   &  1031  \\
Rms residual (ms)                  & 4.3   &  4.7   &  0.78   &  0.46  \\
\hline & \vspace{-3mm} \\
\end{tabular}
\label{tb:para}
\end{table*}

\subsection{Spin-orbit Coupling, $\dot \omega$ and $\dot x$}\label{sec:wex}
\begin{figure} 
\centerline{\psfig{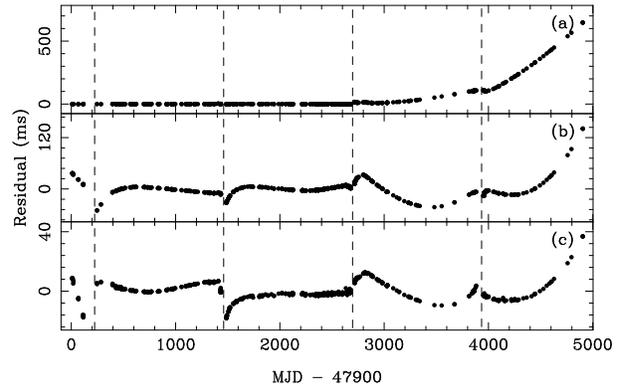}}
\caption{(a) Pre-fit timing residuals using the input parameters from
model 2A of Wex et al. (1998). (b) Post-fit residuals after allowing
all parameters to vary. We obtain
$\dot\omega=0.00029(3)$~deg~yr$^{-1}$ and $\dot
x=61(14)\times10^{-12}$, and the rms residual is 13.1~ms. (c) Post-fit
residuals with the 1997 August glitch included. The rms residual is
4.7~ms.}
\label{fg:mss}
\end{figure}

We next attempted to obtain a timing solution by adding in $\dot
\omega$ and $\dot x$ to account for the spin-orbit coupling.  As a
starting point we used Model 2a from Wex et al. (1998) which used data
up to the end of 1996. Fig.~\ref{fg:mss}a shows this model extended to
the entire data set. Clearly, its predictive power has failed and it
does not track through the 1997 periastron.  The residuals deviate
immediately after the 1997 periastron and become worse again after the
2000 periastron.  We then allowed the orbital parameters, the spin
frequency and its first two derivatives and $\dot \omega$ and $\dot x$
to vary; the resultant fit is shown in Fig.~\ref{fg:mss}b. The
residual has an rms of 13.1~ms and shows significant deviations from
white noise for the 1997 and 2000 orbits. The fitted values of $\dot
\omega$ and $\dot x$ are 0.00029~deg~yr$^{-1}$ and $61\times10^{-12}$
respectively, both having changed sign from the Wex et al. (1998)
solution.  The post-fit residuals after including the glitch are shown
in Fig.~\ref{fg:mss}c. Again, this reduces the rms residual to 4.7~ms,
but leaves systematic variations still unmodelled. Parameters from
this fit are given in the third column of Table~\ref{tb:para}, showing
that both $\dot \omega$ and $\dot x$ remain positive. Fitting for
$\ddot \omega$ and $\ddot x$ changes the sign of $\dot x$ and only
reduces the rms residual to 4.0~ms.

\subsection{Jumps in $\nu$ and $\dot\nu$ at Periastrons}\label{sec:nuj}
\begin{figure} 
\centerline{\psfig{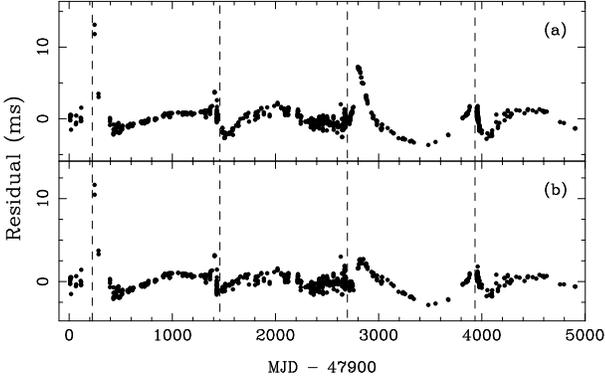}}
\caption{(a) Residuals after fitting for $\nu$, $\dot\nu$, $\ddot\nu$,
the Keplerian orbital parameters and jumps in $\nu$ and $\dot \nu$ at
each periastron, giving an rms residual of 1.4~ms. (b) Post-fit
residuals after adding a glitch at MJD 50691. The rms residual is
0.78~ms.}
\label{fg:jumps}
\end{figure}

Our next attempt at a timing solution involved fitting for frequency
and frequency derivative jumps at each of the four periastron epochs
(in a similar fashion to that done by Manchester et
al. (1995)\nocite{mjl+95} on a very early data set).
Fig.~\ref{fg:jumps}a shows the residuals after fitting for the five
orbital parameters, the rotation frequency and its first two
derivatives and the jumps in $\nu$ and $\dot\nu$ at each
periastron. The rms residual is now only 1.4~ms, and is a significant
improvement on Fig.~\ref{fg:nowx}a and Fig.~\ref{fg:mss}b, albeit at
the expense of having 8 and 6 extra free parameters respectively.
Even this fit, however, does not remove the systematic deviations seen
between the periastron passages.

The magnitude of the jumps in $\nu$ are in the range $10^{-9}$ to
$10^{-8}$s$^{-1}$. For the 1990 and 2000 periastrons they are
negative, while for the 1994 and 1997 periastrons they are
positive. All the jumps in $\dot\nu$ are of negative sign, with
amplitude $\sim 10^{-16}$s$^{-2}$. However, after the glitch is
included, several of the jumps in $\nu$ and $\dot\nu$ change sign:
$\Delta\dot\nu$ at the 1990 and 1994 periastrons become positive,
while $\Delta\nu$ at the 1997 periastron becomes negative. Parameters
of this fit are shown in column four of Table~\ref{tb:para} and the
post-fit residuals are shown in Fig.~\ref{fg:jumps}b.  The rms
residual is now only 0.78~ms, and although some systematic variations
are still seen, the fit is much better than for any previous fit. It
should be noted that the jumps in $\dot\nu$ have absorbed the overall
$\ddot\nu$ term (Table~\ref{tb:para}).

\subsection{Jumps in Orbital Parameters at Periastron}\label{sec:btj}
From the results obtained so far, it seems clear that there are
unmodelled effects still present in the data and that these effects
vary from orbit to orbit but are not accounted for by secular changes
in the orbital parameters. The next logical step is therefore to
test the effect of jumps in the orbital parameters, $\omega$, $x$, $e$ and
$P_{b}$, at each periastron. 

\begin{figure} 
\centerline{\psfig{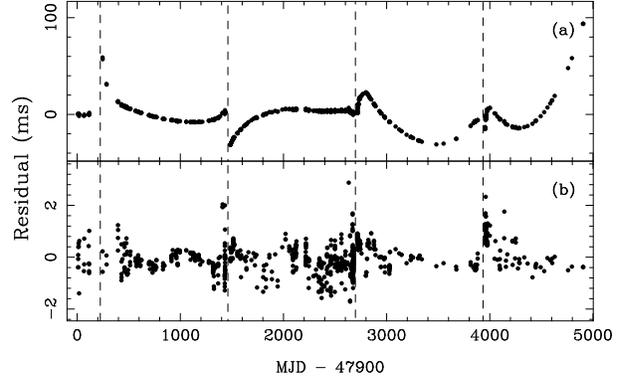}}
\caption{(a) Post-fit residuals for a fit of pulsar frequency and its
first two derivatives, the Keplerian orbital parameters and jumps in
the projected semi-major axis, $x$, at each periastron, giving an
rms residual of 11~ms. (b) Post-fit residuals after the addition of
a glitch at MJD 50691. The rms residual is now only 0.46~ms.}
\label{fg:obj}
\end{figure}

For the data set used by Wex et al. (1998), prior to the 1997
periastron, an excellent fit could be obtained by fitting for the
pulsar frequency, its first two derivatives, the Kepler parameters and
jumps in just $x$ at the 1990 and 1994
periastrons. However, with the full data set and fitting for $\Delta
x$ at the remaining two periastrons, large systematic residuals were
obtained. This fit is shown in 
Fig.~\ref{fg:obj}a which has an rms residual of $\sim$11~ms.
Even allowing for jumps in the other orbital parameters at any
or all of the periastrons did not remove the systematic residual
variations. 

Examination of the post-fit residuals from many of the fits showed a
sharp change of gradient near MJD 50691 (1997 August 30), about 94
days after the 1997 periastron, followed by a roughly exponential
relaxation. This feature is most easily seen in
Fig.~\ref{fg:jumps}a. Remarkably, fitting for this glitch in addition
to jumps in just the projected semi-major axis at each periastron
produced the almost featureless residuals shown in
Fig.~\ref{fg:obj}b. The total frequency change at the time of the
glitch, determined to be MJD 50690.7 $\pm$ 0.7, was
$\Delta\nu_g=(67\pm 1)\times10^{-9}$~s$^{-1}$ of which
$\Delta\nu_d=(22\pm 1)\times10^{-9}$~s$^{-1}$ decayed exponentially
with an assumed timescale of 100 days.
The final rms residual is only 0.46~ms. Only a few points
close to the periastrons are discrepant; these could easily
be due to errors in the DM correction. Parameters for the fit shown in
Fig.~\ref{fg:obj}b are given in column five of
Table~\ref{tb:para}.

\section{Discussion}{\label{sec:dis}}
\subsection{Timing Noise}\label{sec:noise}
PSR B1259$-$63 is a relatively young pulsar with a large
frequency derivative which is likely to have a high level of timing noise
\cite{cd85,antt94}. Arzoumanian et al. (1994) proposed a measure of
the timing noise to be
\begin{equation}
\Delta(t)=\log\left(\frac{|\ddot\nu|}{6\nu} t^3\right)
\end{equation}
where $t$ is the time span over which the pulsar has been observed.
For PSR B1259$-$63 we have $t=10^{8}$~s and for the measured value of
the frequency and its second derivative we obtain
$\Delta_8=-2.1$. This is very close to the value for other pulsars
with similar frequency derivatives, showing that PSR B1259$-$63 is
typical in this respect.


The derived braking index, $n=\nu \ddot\nu / \dot\nu^2$, for PSR
B1259$-$63 is $-36.7$.  Again, this merely reflects the relatively
high level of timing noise in this young pulsar \cite{jg99}. The
absence of systematic residual variations after fitting for steps in
projected semi-major axis at each periastron (Fig.~\ref{fg:obj}b)
suggests that the timing noise is very red; higher-order derivatives
in spin frequency are not significant.

\subsection{Steps in Rotational Parameters}\label{sec:rot}
Fig.~\ref{fg:jumps} shows that jumps in pulsar frequency and frequency
derivative at periastron can account for most of the observed
systematic residuals. As discussed by Manchester et al. (1995), the
most probable cause of frequency jumps at periastron is accretion of
mass and angular momentum from the Be-star disk. The observed
$\Delta\nu$ are of order $10^{-9}$ Hz, requiring accretion of about
$10^{20}$ g at the Alfv\'en radius of $5\times 10^7$ cm. While this
amount of accreted mass is not unreasonable, there are several
problems with this idea. Firstly, the bow shock between the pulsar
wind and the circumstellar wind must lie inside the accretion radius
for accretion to occur. Kaspi et al. (1995) and Tavani \& Arons
(1997)\nocite{ta97} argued that this implies unreasonably high values
for the disk density and/or outflow velocity. Secondly, the observed
$\Delta\nu$ are of different magnitudes and signs at the different
periastrons. While there is evidence from the unpulsed continuum emission
that wind parameters vary from one periastron to the
next \cite{cjmm02}, it would be surprising if the accretion torque were
able to change sign.

A further problem with this explanation for the observed timing
residuals is that frequency jumps alone do not account for the
observed variations; jumps in frequency derivative are also
required. In fact, the $\dot\nu$ jumps are generally more significant
than the jumps in $\nu$ (Table~\ref{tb:para}). The implied changes in
braking torque or moment of inertia are large and physically
improbable. These changes in $\dot\nu$ could result from timing noise,
but the fact that they are concentrated at the periastrons suggests
that this is unlikely.

\subsection{Changes in Orbital Parameters}\label{sec:obj}
It is clear from Section~\ref{sec:wex} that secular changes in
$\omega$ and $x \equiv a\,{\rm sin}\,i$ do not account for the observed
timing behaviour of PSR B1259$-$63, although it worked well for the
first two periastrons. This is reinforced by the fact that the fitted
jumps in $x$ (Section~\ref{sec:btj}) are of different sign at the
different periastrons. This appears to mean that the conclusions drawn
by Wex et al. (1998) about the inclination of the Be-star spin axis to
the orbital plane may no longer be valid. We note that the $\dot x$
measured by Wex et al. (1998) is roughly consistent with the step
change in $x$ at the 1994 periastron described in
Section~\ref{sec:btj}.

It is never-the-less striking that jumps at each periastron in
projected semi-major axis {\it only} are required to satisfactorily
model the data. Limits on changes to the other orbital parameters are
at least an order of magnitude less than the values used in the
simulations of Fig.~\ref{fg:many}.  The systematic residual
variations seen in Fig.~\ref{fg:nowx}b are well modeled by the step
changes in $x$ at each periastron and small adjustments in the other
orbital parameters. If these variations are due to timing noise, the
noise must mimic the effect of changes in the orbital parameters, that
is, be smooth between periastrons and rapid near periastrons, which
seems unlikely.

If the jumps in $x$ are interpreted as changes in
semi-major axis $a$, Kepler's Third Law implies changes in
$P_b$ of order thousands of seconds, orders of magnitude larger than
permitted by the timing solutions (cf. Fig.~\ref{fg:many}). The
observed changes in $x$ must then be interpreted as changes in
inclination angle $i$ of the orbit of order a few arcsec. 

The question then is: what can cause changes in inclination angle
without making significant perturbations to the other orbital
parameters? Furthermore, the mechanism must be able to produce changes
of different sign and magnitude at different periastrons.

There is evidence that the Be-star disk is highly inclined to the
orbital plane both from observations of the pulsed emission
\cite{jwn+01} and the unpulsed emission \cite{jmmc99}, so the velocity
of the disk material would have a significant component perpendicular
to the orbit. In principle, frictional drag could then tilt the
orbit. It is also known that the properties of the wind interaction
vary substantially from orbit to orbit. The problem is that this
effect could only change the orbit inclination with one sign and
Manchester et al. (1995) suggest that the fractional change of orbital
energy due to this effect is $10^{-14}$, too small to account for the
observed changes.

The other possible mechanism appears to be tidal interactions
occurring at each periastron \cite{pp82}. As mentioned in Section 1,
these effects can be greatly enhanced by resonances with modes of
stellar oscillations. Since there is evidently a large angle between
the spin axis and the orbit normal, again it is possible that the
dominant effect could be on the orbit inclination. Different relative
phases of the tides and stellar oscillations could alter the effect on
the orbit from one periastron to the next. Calculations to date
(e.g. Witte \& Savonije 1999) have only considered aligned
systems. However, once again, the predictions are that the effects
would not be significant for PSR B1259$-$63 because of the relatively
large ratio of periastron distance to companion-star radius.

\section{Conclusions}
We have presented an analysis of 13 years of timing data encompassing
four periastron passages of PSR B1259$-$63. Significant changes in
dispersion measure occurring near each periastron were accounted for
in the data analysis. It is clear that secular changes in orbit
parameters due to spin-orbit coupling does not give a good description
of the data. The effect of timing noise is difficult to quantify
because of coupling with the orbital parameters, a problem exacerbated
by the long-duration eclipses which occur at each
periastron. Consequently, we cannot rule out the possibility that
our measured parameters are affected by timing noise. However,
excepting the glitch which occurred near MJD 50691, the fact
that within each orbit the observed residuals show variations with
characteristic timescale of the same order as the orbital period and
significant shorter-term variations only near the periastrons strongly
suggests that changes in orbital parameters at periastron are the
dominant effect. This is reinforced by the excellent fit to the data
by a model containing the MJD 50691 glitch and steps in projected
semi-major axis at each periastron. It is clear that these steps must
result from step changes in the orbit inclination. However, we are
unable to offer any plausible mechanism to account for the observed
changes.

\section*{Acknowledgments}
The Australia Telescope is funded by the Commonwealth of Australia for
operation as a National Facility managed by the CSIRO. We would like
to thank all the many pulsar observers who have contributed to
collecting timing data over the 13 years of this project. NW thanks
NSFCC for support under project 10173020.


\end{document}